\documentclass{mn2e}
\usepackage{epsf,color}
\title[IMF in the Galactic Centre and Starburst Regions]{ The Stellar Mass Spectrum in Warm and Dusty Gas: Deviations from Salpeter in the Galactic Centre and in Circum-Nuclear Starburst Regions}

\author[Klessen, Spaans, \& Jappsen]{Ralf S.\ Klessen$^{1,2}$,
  Marco Spaans$^3$, Anne-Katharina Jappsen$^{2,4}$\\ {$^1$Zentrum f\"ur
  Astronomie der Universit\"at Heidelberg, Institut f\"ur Theoretische Astrophysik,
  Albert-\"Uberle-Str.\ 2,}\\ {~ 69120 Heidelberg, Germany}\\
{$^2$Astrophysikalisches Institut Potsdam, An der Sternwarte 16,
  14482 Potsdam, Germany} \\
{$^3$Kapteyn Astronomical
  Institute, P.O. Box 800, 9700 AV Groningen, The Netherlands}\\
{$^4$Canadian Institute for Theoretical Astrophysics, 
McLennan Physics Labs, 60 St. George Street,}\\{~ University of 
Toronto, Toronto, ON M5S 3H860, Canada}
}

\begin{document}
\def\araa{{\em ARAA}}
\def\aj{{\em AJ}}
\def\apj{{\em ApJ}}
\def\apjl{{\em ApJ}}
\def\apjs{{\em ApJS}}
\def\aap{{\em A\&A}}
\def\apss{{\em Astrophys.\ Space Science}}
\def\baas{{\em Bull.\ Amer.\ Astron.\ Soc.}}
\def\bain{{\em Bull.\ Astron.\ Inst.\ Netherlands}}
\def\fcp{{\em Fund.\ Cosm.\ Phys.}}
\def\jcam{{\em J.\ Comput.\ Appl.\ Math.}}
\def\jcp{{\em J.\ Comput.\ Phys.}}
\def\jfm{{\em J.\ Fluid Mech.}}
\def\mnras{{\em MNRAS}}
\def\nat{{\em Nature}}
\def\pta{{\em Phil.\ Trans.\ A.}}
\def\ptp{{\em Prog.\ Theo.\ Phys.}}
\def\prd{{\em Phys.\ Rev.\ D}}
\def\pre{{\em Phys.\ Rev.\ E}}
\def\prl{{\em Phys.\ Rev.\ Lett.}}
\def\prsa{{\em Proc.\ R.\ Soc.\ London A}}
\def\pasj{{\em Pub.\ Astron.\ Soc.\ Japan}}
\def\pasp{{\em PASP}}
\def\pfl{{\em Phys.\ Fluids}}
\def\ppl{{\em Phys.\ Plasmas}}
\def\rpp{{\em Rep.\ Prog.\ Phys.}}
\def\rmp{{\em Rev.\ Mod.\ Phys.}}
\def\zp{{\em Z.\ Phys.}}
\def\za{{\em Z.\ Astrophys.}}

\date{Received sooner; accepted later}
\pagerange{\pageref{firstpage}--\pageref{lastpage}} \pubyear{2006}
\maketitle

\label{firstpage}

\begin{abstract}
%
%%   Understanding the origin of stellar masses is a key problem in
%%   astrophysics. In the  solar neighborhood, the mass distribution of
%%   stars follows a seemingly universal pattern. In the centre of the
%%   Milky Way, however, there are indications for strong  deviations and
%%   similar may hold for starburst galaxies in the distant universe. We
%%   perform ab-initio hydrodynamic calculations of stellar birth in a
%%   circum-nuclear  starburst region and compare these to the solar
%%   neighborhood. We show that under extreme environmental conditions
%%   the stellar mass spectrum is expected to be  dominated by massive
%%   stars and discuss possible implications for galaxy formation and
%%   evolution.  
%%
  Understanding the origin of stellar masses is a key problem in
  astrophysics. In the solar neighborhood, the mass distribution of
  stars follows a seemingly universal pattern. In the centre of the
  Milky Way, however, there are indications for strong deviations and
  the same may be true for the nuclei of distant starburst galaxies.  Here
  we present the first numerical hydrodynamical calculations of stars
  formed in a molecular region with chemical and thermodynamic
  properties similar to those of warm and dusty circum-nuclear
  starburst regions.  The resulting IMF is top-heavy with a
  peak at $\sim 15$ $M_\odot$, a sharp turn-down below $\sim 7$
  $M_\odot$ and a power-law decline at high masses.  We find a natural
  explanation for our results in terms of the temperature dependence
  of the Jeans mass, with collapse occuring at a temperature of $\sim
  100$ K and an H$_2$ density of a few $\times 10^5$ cm$^{-3}$, and
  discuss possible implications for galaxy formation and evolution.
\end{abstract}


\section{Introduction}

Identifying the physical processes that determine the masses
of stars and their statistical distribution, the initial mass function
(IMF), is a fundamental problem in star-formation research. It is
central to much of modern astrophysics, with implications ranging from
cosmic re-ionisation and the formation of the first galaxies, over the
evolution and structure of our own Milky Way, down to the build-up of
planets and planetary systems.
%  which have been discovered in abundance in the past decade. 

Near the Sun the number density of stars as a function
of mass has a peak at a characteristic stellar mass of a few tenths of
a solar mass, below which it declines steeply, and for masses above
one solar mass it follows a power-law with an exponent $dN/d{\log}m
\propto m^{-1.3}$. Within a radius of several kpc %from the Sun
this distribution shows surprisingly little variation (Salpeter 1955;
Scalo 1998; Kroupa 2001; Kroupa 2002; Chabrier 2003).  This has
prompted the suggestion that the distribution of stellar masses at
birth is a truly universal function, which often is referred to as the
Salpeter IMF, although note that the original Salpeter (1955) estimate
was a pure power-law fit without characteristic mass scale.

On the other hand, there is increasing evidence that the IMF close to
the centre of our Milky Way (Stolte et al.\ 2002, 2005, Nayakshin \&
Sunyaev 2005, Paumard et al.\ 2006) and the neighboring Andromeda
galaxy (Bender et al.\ 2005) is dominated by massive rather than
low-mass stars. For the circum-nuclear starburst regions in more
distant galaxies, very similar IMF deviations are subject to
continuing debate (e.g., Scalo
1990, Elmegreen 2005). %% If true, this would have important consequences
%% for our understanding of the early universe and for galaxy evolution
%% from high redshifts to present times (10). 
However, no conclusion has yet been reached, and it appears timely to
examine the problem from a theoretical point of view.

We approach the problem by means of self-consistent hydrodynamical
calculations of fragmentation and star formation in interstellar gas
where chemical and thermodynamical properties are described by a
realistic equation of state (EOS). We focus on the most
extreme environmental
conditions such as occur in the nuclear regions of massive
star-forming spiral galaxies. There the inferred dust and gas
temperatures, gas densities and star formation rates typically exceed
the solar-neighborhood values by factors of 3, 10 and $\ge 100$,
respectively (e.g.\ Ott et al.\ 2005; Israel 2005; Aalto et al.\ 2002;
Spinoglio et al.\ 2002). Consequently, it has long been speculated
that such conditions lead to deviations from the Salpeter IMF (e.g.,
Scalo 1990, Elmegreen 2005). 
%Our study shows that this is indeed expected to be the case.

\begin{figure*}          
\centerline{%
\epsfxsize=0.90\textwidth\epsffile{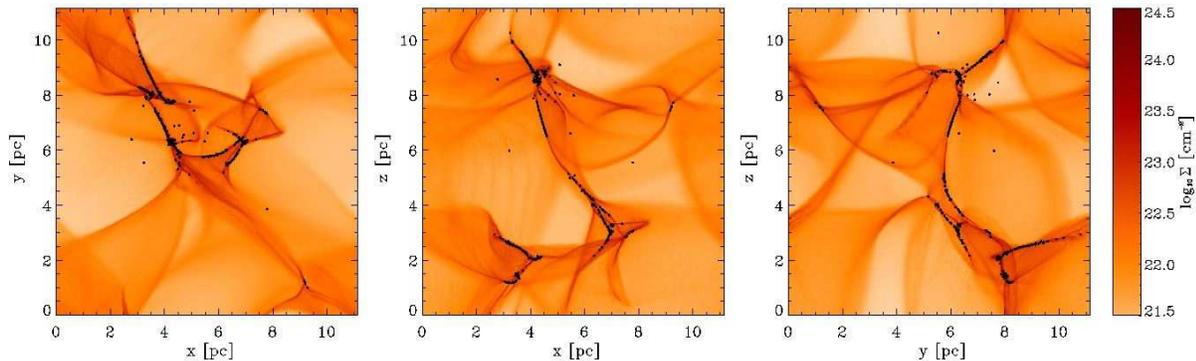}
}
\vspace*{0.4cm}
\caption{Column density distribution of the gas  projected along
  the three principal axes of the system after 3.5 million years of
  evolution  when 15\% of the gas is converted into protostars
  (sink particles). Their location is indicated by black dots.} 
\label{fig:2D-plot}
\end{figure*}

\section{Model}

Stars and star clusters form through the interplay between
self-gravity on the one side and turbulence, magnetic fields, and
thermal pressure on the other
%, so-called gravoturbulent fragmentation
(for recent reviews see Larson 2003; Mac~Low \& Klessen 2004;
Ballesteros-Paredes et al.\ 2006). The supersonic turbulence
ubiquitously observed in interstellar gas clouds can create strong
density fluctuations with gravity taking over in the densest and most
massive regions. Collapse sets in to build up stars and star
clusters. Turbulence plays a dual role. On global scales it
provides support, 
%and counterbalances gravity,
on local scales it provokes collapse. Stellar birth is thus intimately
linked to the dynamic behavior of the parental gas cloud, which
governs when and where star formation sets in (as illustrated in
Figure \ref{fig:2D-plot}).

The chemical and thermodynamic properties of interstellar clouds play
a key role in this process. In particular, the value of the polytropic
exponent $\gamma$, when adopting an EOS of the
form $P\propto\rho^\gamma$, strongly influences the compressibility of
density condensations as well as the temperature of the gas. The EOS
thus determines the amount of clump fragmentation, and so directly
relates to the IMF (V\'azquez-Semadeni et al.\ 1996) with values of
$\gamma$ larger than unity leading to little fragmentation and high
mass cores (Li, Klessen, \& Mac~Low 2003; Jappsen et al.\ 2005). The
stiffness of the EOS in turn depends strongly on the ambient
metallicity, density and infrared background radiation field produced
by warm dust grains. The EOS thus varies considerably in different
galactic environments (see Spaans \& Silk 2000, 2005 for a detailed
account).

For the circum-nuclear starburst regions that are the subject here we
assume a cosmic ray ionisation rate of $3\times 10^{-15}$ s$^{-1}$,
solar relative abundances (Asplund et al.\ 2004; Jenkins 2004) and an
overall metallicity of two times solar (Barthel 2005).  A velocity
dispersion $\Delta V_{\rm tur}=5$ km/s is adopted to take the larger
input of kinetic energy (e.g.\ through supernovae) into account. The
dust temperature inside the model clouds is set by a fiducial
background star formation rate of $100\,$M$_\odot$ yr$^{-1}$/kpc$^{2}$
which causes dust grains to be at temperatures of about $T_d=30 -
90\,$K, depending on the amount of shielding. Gas temperatures range
from $T_g = 40 - 140\,$K, over a density range of $10^4 -
10^7\,$cm$^{-3}$. These values are consistent with gas and dust
temperatures determined for circum-nuclear starburst regions (Klaas et
al.\ 1997; Aalto et al.\ 2002; Spinoglio et al.\ 2002; Ott et al.\ 
2005; Israel 2005).

\begin{figure}        
\centerline{\epsfxsize=0.45\textwidth\epsffile{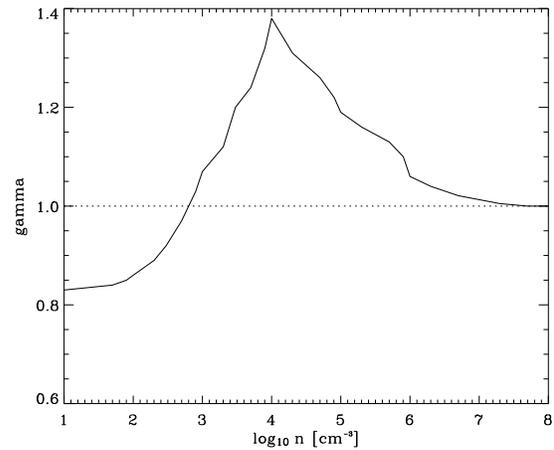}}
%\centerline{\epsfxsize=0.40\textwidth\epsffile{spaans-EOS-3.ps}}
\caption{Starburst EOS adopted from Spaans \& Silk (2005).}
\label{fig:starburst-EOS}
\end{figure}

Figure \ref{fig:starburst-EOS} shows the resulting polytropic exponent
as a function of density. The main feature is the $\gamma > 1$ peak
around $n = 10^4\ $cm$^3$. This peak implies that the gas warms up as
it is compressed and it is caused mainly by strong photon trapping in
opaque H$_2$O and CO lines in the metal-rich nuclear gas. That is, the
large optical depth in the cooling lines suppresses the cooling
efficiency. Also, warm dust ($T > 40\,$K), heated by the ambient
stars, causes H$_2$O collisional de-excitation heating through
far-infrared pumping (Takahashi, Hollenbach \& Silk, 1983;
Spaans \& Silk 2005), which adds to the gas-dust
heating.  Cosmic-ray heating rate is elevated by a high supernova
rate, as expected for nuclear starburst regions (Bradford et al.\ 
2003).

%% shows the resulting dependence of the polytropic exponent on
%% density.  Its main feature is the $\gamma >1$ peak around a few
%% $\times 10^4$ cm$^{-3}$. This peak implies that the gas warms up as it
%% is compressed and it is caused mainly by strong photon trapping in
%% opaque H$_2$O and CO lines (Spaans \& Silk 2000, 2005).  Also, warm
%% dust ($>40$ K) causes H$_2$O collisional de-excitation heating through
%% far-infrared pumping (Takahashi, Hollenbach \& Silk 1983; Spaans \&
%% Silk 2005), which adds to the gas-dust and cosmic-ray heating rates at
%% these densities.

Adopting this EOS we follow the dynamical evolution of the
star-forming gas using smoothed particle hydrodynamics (SPH).  This is
a Lagrangian method to solve the equations of hydrodynamics, where the
fluid is represented by an ensemble of particles, and flow quantities
are obtained by averaging over an appropriate subset of SPH particles
(Monaghan 2005). The method is able to resolve high density
contrasts as particles are free to move, and so the particle
concentration increases naturally in high-density regions.  The
performance and convergence properties of SPH are well understood and
tested against analytic models and other numerical schemes in the
context of astrophysical flows (see, e.g., Mac~Low et al.\ 1998;
Lombardi et al.\ 1999; Klessen et al.\ 2000; O'Shea et al.\ 2005;
Ballesteros-Paredes et al.\ 2006). Artificial fragmentation can be
ruled out, as long as the mass within one smoothing volume remains
less than half the critical mass for gravitational collapse (Bate \&
Burkert 1997; Hubber, Goodwin, \& Whitworth 2005).  We use the
publically available parallel code GADGET (Springel et al.\ 2001). It
is modified to replace high-density cores with sink particles (Bate,
Bonnell, \& Price 1995) that can accrete gas from their surroundings
while keeping track of mass and momentum.  This enables us to follow
the dynamic evolution of the system over many local free-fall
timescales.  We identify sink particles as the direct progenitors of
individual stars.  For a more detailed account of the method and a
discussion of its convergence properties we refer the reader to
Klessen et al.\ (2000) and Jappsen et al.\ (2005).

\begin{figure*}          
\centerline{\epsfxsize=0.45\textwidth\epsffile{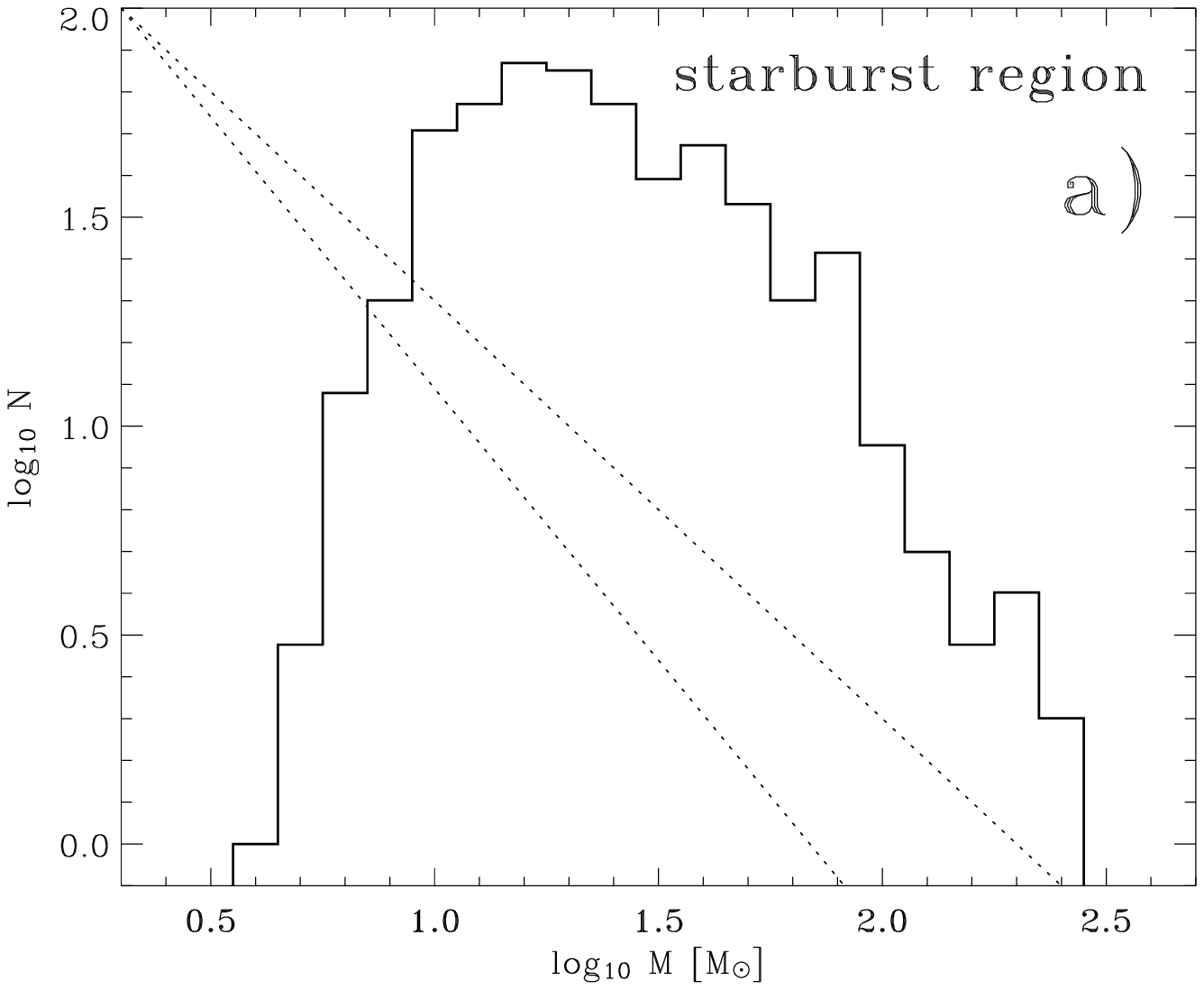}\epsfxsize=0.45\textwidth\epsffile{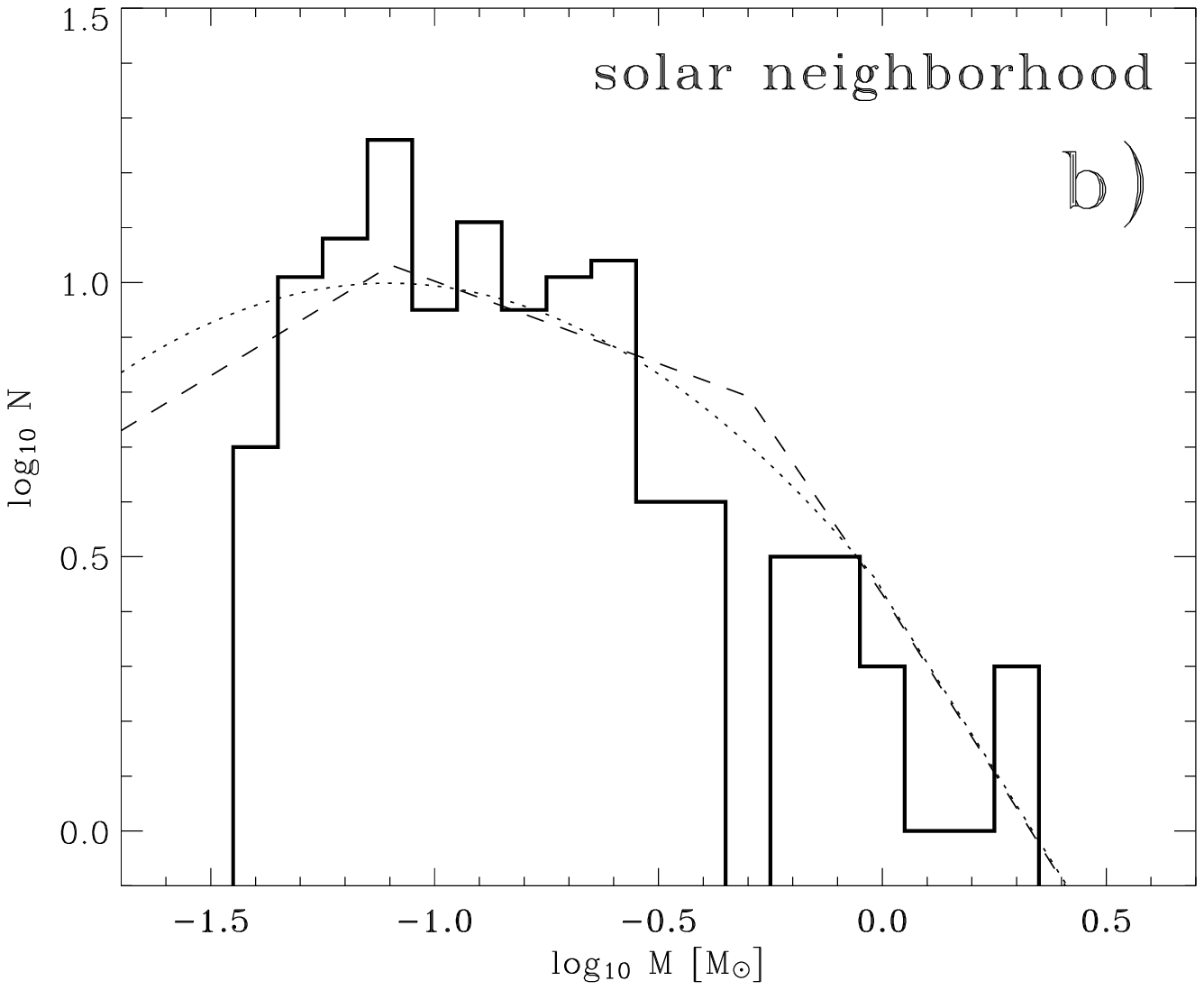}}
\caption{
  {\em (a)} Mass spectrum of gravitational condensations in the
  starburst calculation at a time when 15\% of the gas is converted
  into collapsed objects (which we identify as direct progenitors of
  individual stars). To guide the eye, we indicate a slope -1.0 and
  the Salpeter slope -1.3 with dotted lines. The mass function in our
  simulated starburst environment shows a broad peak in the range
  $10\,$M$_{\odot}-25\,$M$_{\odot}$ and falls off for larger masses.
  It is thus top-heavy compared to the IMF in the solar neighborhood
  (Salpeter 1955; Kroupa 2002; Chabrier 2003).  {\em (b)} Mass
  spectrum of collapsed objects in a calculation focusing on nearby
  molecular clouds (see Jappsen et al.\ 2005).  It agrees well with
  the IMF in the solar vicinity. For comparison we overplot the
  functional forms proposed by Kroupa (2002) with dashed lines and by
  Chabrier (2003) with dotted lines. Our two calculations differ
  mainly in the adopted EOS, i.e.\ in the chemical and thermodynamic
  state of the star forming gas, other parameters are comparable.
%%
%%   Mass spectrum of protostellar objects in a simulation
%%   focusing on a starburst environment. The size of the considered
%%   molecular cloud region is $11.14\,$pc and it contains
%%   $80\,000\,$M$_{\odot}$. The system is depicted at a time when 15\% of
%%   the gas is converted to stars. The mean mass of the distribution
%%   lies at 15 $M_\odot$. The nominal resolution limit is $1\,$M$_{\odot}$.
}
\label{fig:starburst-IMF}
\end{figure*}

We focus on a cubic volume of 11.2$\,$pc in size, which contains
$80,000\,$M$_{\odot}$ of gas and has an initial mean particle density
$n = 10^3\,$cm$^{_3}$ at a temperature of $21\,$K. Above the
characteristic density $n = 10^4\,$cm$^{-3}$ where $\gamma$ is at a
maximum, the temperature quickly reaches values of $\sim 100\,$K. This
set-up is chosen to describe the typical environment within the
central regions of an actively star-forming galaxy such as our own
Milky Way or NGC$\,$253. In such galaxies, high-density gas with $n >
10^5\,$cm$^{-3}$, as traced by HCN, typically has a filamentary
structure with very low filling factor, while the bulk of the gas is
at $n \approx 10^3\,$cm$^{-3}$ (Morris \& Serabyn 1996; H\"uttemeister
et al.\ 1993; Israel \& Baas 2003), exactly as found at the end of our
simulation (see Figure \ref{fig:2D-plot}). We stop the calculation at
a star formation efficiency SFE $\approx 15$\%, when roughly 1/6
of the total gas mass has turned into gravitationally collapsed condensations
(i.e.\ sink particles, which we identify as direct progenitors of
individual young stars).
%% For typical molecular clouds in the solar vicinity less
%% than a few percent of their mass takes part in star formation (Myers
%% et al.\ 1986).  However, in circum-nuclear starburst regions this star
%% formation efficiency (SFE) can be higher by as much as an order of
%% magnitude (Mooney \& Solomon 1988).
%%
Throughout the simulation we drive turbulence continuously on large
scales, with wave numbers $k$ in the range $1 \le k \le 2$ (see
Mac~Low 1999) to yield a constant turbulent Mach number
$\mathcal{M}_{\mathrm{rms}}\approx 5$. The particle number is
$N=8\,000\,000$. 
This is thus one of the
highest-resolution star-formation calculations done with SPH,
with a total CPU time of $8 \times 10^4$ hours.
%%  The star formation
%% efficiency, i.e.\ the total fraction of gas turned into stars at the
%% end of the simulation, is about 15\%.  
The critical density for sink particle formation is $n_c =
10^7\,$cm$^{-3}$, with a sink particle radius of 0.015$\,$pc.  The
mass of individual SPH particles is $m=0.01\,$M$_{\odot}$, which is
sufficient to resolve the minimum Jeans mass in the system $M_{\rm J}
\approx 1.5\,$M$_{\odot}$. Except for the EOS and the particle number,
the numerical set-up is identical to the study by Jappsen et al.\ 
(2005). We have performed a second run for a region of 5.7$\,$pc with
4 times less mass, eight times fewer particles and a sink particle
radius of 0.02 pc that has reached a SFE $\sim 36$\%.
%% {\sl 
%% For SFE $>10$\%, the resulting
%% mass spectrum in the two regions are statistically indistinguishable
%% and no longer change with time. We are thus allowed to consider the
%% results of the two simulations in combination.
%% }

\section{Result and Physical Interpretation}

We find that in the considered star-forming region, the mass spectrum
of collapsed objects is biased towards high masses. The resulting IMF
has a broad peak at $\sim 15\,$M$_{\odot}$ followed by an approximate
power-law fall-off with a slope in the range -1.0 to -1.3.
Furthermore, there is a clear deficit of stars below $7\,$M$_{\odot}$.
This is illustrated in Figure \ref{fig:starburst-IMF}a.  We contrast
this finding with the result from a simulation appropriate for the
physical conditions in star forming regions near the Sun (from Jappsen
et al.\ 2005), where $\gamma$ changes from 0.7 to 1.1 at an H$_2$
density of a few $\times 10^5\,$cm$^{-3}$. As expected, Figure
\ref{fig:starburst-IMF}b shows a mass spectrum that is very similar to
the IMF in the solar neighborhood (Kroupa 2002; Chabrier 2003). These
striking differences are caused by the very disparate chemical and
thermodynamic state of the star forming gas in the two simulations,
since all other parameters are very similar. Our results thus support
the hypothesis that for extreme environmental conditions as inferred
for the centres of most spiral galaxies or more general for
IR-luminous circum-nuclear starburst regions the IMF is indeed
expected to be top-heavy.

%%  Note,
%% that the strong shear motions close to the Galactic centre may be able
%% to mimic the EOS effects discussed here, as shear adds stability and
%% thus requires larger Jeans masses for collapse to occur. However, the
%% Arches cluster is bound, i.e. this shear field may not play a dominant
%% role in the inner parts of the cluster.

There is a natural explanation for our results in terms of the
temperature dependence of the Jeans mass $M_{\rm J}$.  Compared to a
mean temperature of $10\,$K for dense molecular gas in the Milky Way,
gravitationally collapsing gas in our simulations has a temperature of
$\sim 100\,$K and an H$_2$ density of a few $\times 10^5$ cm$^{-3}$.
As the critical mass for gravitational collapse scales as $M_{\rm J}
\propto T^{1.5}$, this boosts $M_{\rm J}$ from $0.3\,$M$_\odot$ at
$10\,$K to about 10$\,$M$_\odot$ at $100\,$K (see also Klessen \&
Burkert 2000, Bonnell, Clarke, \& Bate 2006). This temperature may seem
high, but is quite consistent with molecular cloud observations in the
Galactic centre (e.g. H\"uttemeister et al.\ 1993) or with
high-density ($n>10^4\,$cm$^{-3}$) NH$_3$ data in the
starburst centre of NGC253 (Ott et al.\ 2005). We also note, that this
Jeans mass scaling argument is supported by recent observations in
more nearby high-mass star-forming regions. For example, in M17 at a
distance of 1.6$\,$kpc from the Sun, the mass spectrum of prestellar
cores, which are the direct progenitors of individual stars, peaks at
at $\sim 4\,$M$_{\odot}$ at an ambient temperature of $30\,$K (Reid \&
Wilson 2006). This is well above the corresponding peak in low-mass
star-forming regions (e.g.\ Motte, Andr{\'e}, \& Neri 1998). 
%% In principle,
%% the formation of multiple stellar systems (28) and protostellar
%% feedback (29), which our numerical method cannot resolve, may diminish
%% the final stellar masses. Still, even in the extreme case that half of
%% the collapsing mass is redistributed or removed, deviations from the
%% standard IMF will persist.

\section{Discussion}

Our mass spectrum is in good agreement with the IMF estimates in the
Galactic centre by Stolte et al.\ (2002, 2005), Nayakshin et al. (2005), and
Paumard et al.\ (2006).  For example, Stolte et al.\ (2002, 2005) find for
the Arches cluster a clear deficit of stars below $7\,$M$_\odot$.  This is
consistent with our result in the sense that the ambient
densities and temperatures found in the Galactic centre are similarly
elevated (Helfer \& Blitz 1996) as in the circum-nuclear starburst
environment we consider. We stress that the turn-down in our model IMF
at masses below 10$\,$M$_\odot$
is a direct consequence of the stiff EOS  for densities $n$ above a
few$\times 10^3\,$cm$^{-3}$ through the Jeans mass 
temperature dependence, and is not caused by resolution effects.  Our
two simulations resolve masses down to $\sim 2\,$M$_\odot$ and $\sim
1\,$M$_\odot$, respectively, and our least massive stars (i.e.\ sink
particles) are well above this limit. Rather, the effective Jeans mass
at $T\sim 100$K and densities of $\sim 10^5-10^6$ cm$^{-3}$ prevent
the formation of low-mass stars.

When interpreting our simulation results, there are several caveats
that need to be kept in mind.  First, our numerical model does not
include shear. Strong shear motions may mimic the EOS effects
discussed here, as shear adds stability and thus requires larger Jeans
masses for collapse to occur. However, the Arches cluster is bound.
Thus the Galactic centre shear field cannot play a dominant role in
the inner parts of the cluster.  Second, our numerical model does not
take the effects of magnetic fields into account, which may be of
considerable strength in the Galactic centre (Yusef-Zadeh \& Morris
1987, but also see Roy 2004 for lower estimates). However, even if
there is a rough equipartition between  kinetic and magnetic
energy, the chemical and thermodynamic properties of the
gas are not strongly affected. Our results will still hold at
least qualitatively, in the sense that an extreme environment leads to
deviatiations from the standard Salpeter IMF. Third, the use of sink
particles does not permit us to resolve close binary systems. Massive
stars in the solar vicinity are almost always members of a binary or
higher-order multiple stellar system (e.g.\ Vanbeveren et al.\ 1998).
If this trend holds also for starburst environments, then the peak of
the stellar IMF will lie below the value reported here. For instance,
if each unresolved sink particle in our calculation separates into a
binary star, in a statistical sense our mass spectrum needs to be
shifted to lower masses by a factor of 0.5. Finally, protostellar
feedback may locally affect the accretion onto individual protostars.
In this case the mass content of the sink particle may only poorly
reflect the mass that ends up in a star.  However, even in the extreme
case that half the mass is removed by feedback during collapse (for
estimates, see Yorke \& Sonnhalter 2002; Krumholz, McKee, \& Klein
2005), deviations from the standard IMF will still persist.

For typical molecular clouds in the Milky less than a few percent of
their mass takes part in star formation (e.g.\ Myers et al.\ 1986) and
this fraction goes up by a factor of a few for cluster-forming cores
(e.g.\ Lada \& Lada 2003). A number of observations (Paglione et al.
1997; Mooney \& Solomon 1988) indicate that starburst systems like
NGC253 and M82, and luminous infrared galaxies in general, have a
larger fraction of their interstellar gas mass at high densities (Gao
\& Solomon 2004). Consequently, their SFE's are up by as much as an
order of magnitude. Our simulations cover this range and the
statistics of our mass spectra do not change above a SFE $\sim 10$\% in
both runs. Hence, the precise SFE that pertains to a starburst
environment does not influence our results as long as it is larger
than 10\%.

The computed star formation rate (SFR), defined as the change in mass
with time of the sink particles, is typically $860\, M_\odot$
yr$^{-1}/$kpc$^2$ for a SFE $>10$\% and when normalised to a surface area of
$1\,$kpc$^2$, which is roughly the size scale of the nuclear region
inside a starburst galaxy. This number lies well within the fiducial
range of $50 - 1000\,M_\odot$yr$^{-1}/$kpc$^2$ inferred for most
starburst systems (e.g.\ Kennicutt 1998, Scoville \& Wilson 2004).

When turning to distant starburst galaxies in the early universe, the
low-mass cut-off at $7\ $M$_{\odot}$ seen in the simulated local
starburst region seems at first glance difficult to reconcile with the
mass-to-light ratio and the stellar population synthesis models
inferred from global observations (Kaufmann et al.\ 2003).
However, we emphasise again, that we are focusing on an extreme case
and on a clearly localised, isolated region only. In reality these
extreme (warm and dusty) environmental conditions will not apply to
all regions inside a starburst galaxy.  There will be pockets of
colder gas with different ($\gamma < 1$) EOS that are less exposed to
radiation (Spaans \& Silk 2000) and that behave like Galactic
star-forming regions. Under these conditions the studies by Jappsen et
al.\ (2005) and Larson (2005) indicate that a normal, Salpeter-like
IMF results.  This also suggests that the relative contribution of the
extreme IMF found in this work can be connected directly to the
observations. The fraction of molecular gas at densities $> 10^4$
cm$^{-3}$ that enjoys temperatures larger than $50\,$K should be a
strong indicator of deviations from a Salpeter IMF.  Future work will
address the issue of stellar population matching and will compare our
results with observed M/L ratios and warm, high density gas mass
estimates.

\label{lastpage}

\end{document}